\documentclass[showpacs,aps,graphicx,twocolumn]{revtex4}%
\usepackage{graphicx}

\begin{document}
\title{Quantum secure direct communication network with Einstein-Podolsky-Rosen pairs}
\author{ Fu-Guo Deng,$^{1,2,3}$\footnote{E-mail addresses: fgdeng@bnu.edu.cn}
Xi-Han Li,$^{1,2}$ Chun-Yan Li,$^{1,2}$ Ping Zhou,$^{1,2}$ and
Hong-Yu Zhou$^{1,2,3}$}
\address{$^1$ The Key Laboratory of Beam Technology and Material
Modification of Ministry of Education, Beijing Normal University,
Beijing 100875, People's Republic of China\\
$^2$ Institute of Low Energy Nuclear Physics, and Department of
Material Science and Engineering, Beijing Normal University,
Beijing 100875, People's Republic of China\\
$^3$ Beijing Radiation Center, Beijing 100875,  People's Republic
of China}
\date{\today }

\begin{abstract}
We discuss the four requirements for a real point-to-point quantum
secure direct communication (QSDC) first, and then present two
efficient QSDC network schemes with an $N$ ordered
Einstein-Podolsky-Rosen pairs. Any one of the authorized users can
communicate another one on the network securely and directly.
\end{abstract}
 \maketitle

\section{INTRODUCTION}

The combination of the features of quantum systems with information
has produced many interesting and important developments in the
field of the transmission and the processing of information. Quantum
key distribution (QKD), an important application of quantum
mechanics supplies a secure way for creating a private key between
two remote parties, the sender, Bob and the receiver, Carol. The
noncloning theorem \cite{noncloning} forbids a vicious eavesdropper,
Eve to copy perfectly the quantum signal transmitted through the
quantum line, and her action will inevitably disturb the quantum
system and leave a trick in the results. Bob and Carol can find out
Eve by comparing some of the results chosen randomly and analyzing
its error rate. Combined with a private key, secret message can be
transmitted securely with one-time-pad crypto-system. QKD has
progressed quickly
\cite{bb84,ekert91,bbm92,b92,gisin,duan,longqkd,CORE,BidQKD,ABC,Hwang}
since Bennett and Brassard proposed the standard BB84 QKD protocol
\cite{bb84} in 1984. The reason may be that the modern technology
allows QKD to be demonstrated in laboratory \cite{gisin} and
practical applications can be achieved in the future.

Recently, a novel branch of quantum communication, quantum secure
direct communication (QSDC) was proposed and actively pursued by
some groups
\cite{beige,bf,two-step,QOTP,Wangc,Wangc2,cai,yan,zhangzj,Gao,caiA,Nguyen,mancpl,Lucamaini}.
With QSDC Bob and Carol can exchange the secret message directly
without generating a private key in advance and then encrypting the
message, which is different to QKD. In 2002, Beige et al.
\cite{beige} presented a QSDC protocol in which the message can be
read out after the transmission of an additional classical
information for each qubit \cite{bf,two-step,QOTP}, similar to a QKD
scheme as each bit of key can represent one bit of secret message
with an additional classical information, i.e., retaining or
flipping the bit value in the key according to the secret message
\cite{two-step}. The same case takes place in Refs.
\cite{yan,zhangzj,Gao}. In 2002, Bostr\"{o}m and Felbinger proposed
a ping-pong QSDC following some ideas in quantum dense coding
\cite{bw} with an Einstein-Podolsky-Rosen (EPR) pair. The authors
have claimed that it is secure for generating a private key and
quasi-secure for direct communication as it will leak some of the
secret message in a noise channel \cite{bf}. W$\acute{o}$jcik and
Zhang et al. pointed out that the ping-pong protocol is insecure for
direct communication if there are losses in a practical quantum
channel \cite{attack1,attack2}. Also, the ping-pong protocol
\cite{bf} can be attacked without eavesdropping
\cite{attack4,attack5}.  Cai and Li \cite{cai} modified the
ping-pong protocol for transmitting the secret message directly by
replacing the entangled photons with a single photon in a mixed
state, similar to the Bennett 1992 QKD \cite{b92}. Meanwhile, Deng
et al. put forward a two-step QSDC protocol \cite{two-step} with EPR
pairs transmitted in block and another one based on a sequence of
polarized single photons \cite{QOTP}. Wang et al. \cite{Wangc}
introduced a QSDC protocol with high-dimension quantum superdense
coding. The good nature of the QSDC schemes
\cite{two-step,QOTP,Wangc,Wangc2} with quantum data block is that
the parties can perform quantum privacy amplification
\cite{ep,deutschqpa,QPA} on the unknown states for improving their
security in a noise channel. In Ref. \cite{caiA}, Cai and Li
designed a protocol for improving the capacity of the ping-pong QSDC
protocol \cite{bf} with the same way for eavesdropping check as that
in Ref. \cite{two-step}. However, it is not unconditionally secure
as the analysis of eavesdropping check depends on the feature of
statistics for which a lot of samples should be chosen randomly and
measured. Recently, Lucamarini et al. \cite{Lucamaini} introduced a
QSDC protocol for both  direct communication and  creating a private
key with the same ideas in the Refs. \cite{BidQKD,QOTP}. It is
secure for QKD, same as Ref. \cite{BidQKD}, but it is just
quasi-secure for direct communication, similar to the QSDC protocol
in Ref. \cite{caiA}.

By far, there are many QKD network schemes
\cite{Phoenix,Townsend,Biham,MUQKDguo,DLMXL,LZWD} in which one user
can communicate any other one on the network, but not a QSDC network
scheme even though there are some point-to-point QSDC schemes
\cite{beige,bf,two-step,QOTP,Wangc,Wangc2,cai,yan,zhangzj,Gao,caiA,Nguyen,mancpl,Lucamaini}
existing. Moreover, almost all of the existing QSDC point-to-point
schemes cannot be used directly to accomplish the task in a QSDC
network as a distrustful server can steal some information without
being detected. In this paper, we will introduce two QSDC network
schemes with an ordered $N$ EPR photon pairs. One authorized user
can communicate any one on the network securely with the capability
of single-photon measurements in the first scheme.  As almost all
the instances are useful and each EPR photon pair can carry two bits
of information, the intrinsic efficiency for qubits and the source
capacity are both high. In the second scheme, the users can exploit
entanglement swapping to transmit the secret message securely after
they set up the quantum channel, the EPR pairs shared. As the qubits
encoded by the sender do not suffer from the noise of the quantum
line again, its security may be higher than the first one. On the
other hand, it has only half the source capacity of the first
scheme. Also, the four requirements for a real secure point-to-point
quantum direct communication scheme are discussed in detail.

\section{The requirements for a real QSDC}

From the way for transmitting the quantum data and analyzing the
security of the quantum channel, all existing QSDC schemes can
attributed to one of the two types, the one in which the quantum
signal is transmitted in a stream (stream-QSDC) and the other in a
quantum data block (QDB-QSDC). The feature of stream-QSDC
\cite{bf,cai,caiA,Nguyen,mancpl,Lucamaini} is that Bob and Carol
choose randomly the eavesdropping check mode or the message-coding
mode with two asymmetric probabilities for the quantum signal
transmitted in a stream (one photon in each round). In the check
mode, Bob and Carol obtain a sample for eavesdropping check by means
that they choose one or two sets of measuring bases (MBs) to measure
it \cite{bf,cai,caiA,Nguyen,mancpl,Lucamaini}. When they choose the
message mode, they  encode the secret message on the quantum state
directly. In a word, the security check and the encoding of the
secret message are done concurrently in stream-QSDC protocols
\cite{bf,cai,caiA,Nguyen,mancpl,Lucamaini}. The property of QDB-QSDC
\cite{two-step,QOTP,Wangc,Wangc2,yan,zhangzj,Gao} is that the
quantum signal is transmitted in a quantum data block. That is, Bob
and Carol have to transmit a sequence of quantum states and check
its security before Bob encodes the secret message on them. In
brief, the encoding of the secret messages is done only after the
confirmation of the security of the quantum channel
\cite{two-step,QOTP,Wangc,Wangc2} is accomplished.

In essence, the security of quantum communication bases on the two
principles: (1). one is the properties of quantum states, such as
the uncertainty principle (no-cloning theorem), quantum
correlations, non-locality, and so on; (2). the other is the
analysis for quantum error rate based on the theories in
statistics. The first principle ensures that Eve cannot copy the
quantum states freely as her action will inevitably perturb the
quantum systems, which will introduce some errors in the results.
The second one is used to check the security of the quantum
channel after Bob and Carol transmit the sufficient quantum
states. The check for eavesdropping is valid only when Bob and
Carol can sample sufficiently enough instances from results
transmitted. That is, the message may be secure only when they are
obtained after checking eavesdropping.

For QKD Bob and Carol can choose randomly one of two MBs for the
quantum states transmitted in one by one as the analysis of the
eavesdropping check is just a postprocessing. The security of QKD
requires them to determine whether there is an eavesdropper
monitoring the quantum channel. The case in QSDC is different to QKD
as the two parties cannot abandon the secret message transmitted. In
brief, a real point-to-point QSDC protocol should satisfy the four
requirements: (1) the secret message can be read out by the receiver
directly after the quantum states are transmitted through a quantum
channel, and there is no additional classical information exchanged
by the sender and the receiver in principle except for those for
checking eavesdropping and estimating the error rate. (2) the
eavesdropper, Eve cannot obtain an useful information about the
secret message no matter what she does; In another word, she can
only get a random result for the message with her eavesdropping on
the quantum signal. (3) the two legitimate users can detect Eve
before they encode the secret message on the quantum states. (4) the
quantum states are transmitted in a quantum data block. The last one
is not necessary for QKD as the two authorized users just distribute
a key which does not include the information about the secret
message in this time and can be abandoned if they find out Eve
monitoring the quantum channel. QSDC is used for directly
communicating the secret message which cannot be discarded. The
security of quantum communication depends on the analysis for
quantum error rate based on the theories of statistics in which many
samples are chosen randomly for its accuracy. In this way, the
quantum states should be transmitted in a quantum data block in a
QSDC.

From the view of security, the QDB-QSDC protocols
\cite{two-step,QOTP,Wangc,Wangc2} are secure with some other quantum
techniques, such as quantum privacy amplification
\cite{deutschqpa,QPA}, in a noise quantum channel. The stream-QSDC
protocols \cite{bf,cai,caiA} are just quasi-secure as the authorized
users cannot take a quantum privacy amplification on the quantum
states transmitted one by one, which is in principle different to
the QDB-QSDC protocols.

\section{QSDC network with EPR pairs}
\subsection{Bidirectional QSDC network}

An EPR pair can be in one of the four Bell states \cite{book},
\begin{eqnarray}
\vert \psi ^{-}\rangle_{BC} &=&\frac{1}{\sqrt{2}}(\vert 0\rangle
_{B}\vert 1\rangle _{C} - \vert 1\rangle _{B}\vert 0\rangle
_{C})\\
\vert \psi ^{+}\rangle_{BC} &=&\frac{1}{\sqrt{2}}(\vert 0\rangle
_{B}\vert 1\rangle _{C} + \vert 1\rangle _{B}\vert 0\rangle _{C})
\label{EPR2}
\\
\vert \phi ^{-}\rangle_{BC} &=&\frac{1}{\sqrt{2}}(\vert 0\rangle
_{B}\vert 0\rangle _{C} - \vert 1\rangle _{B}\vert 1\rangle _{C})
\\
\vert \phi ^{+}\rangle_{BC} &=&\frac{1}{\sqrt{2}}(\vert 0\rangle
_{B}\vert 0\rangle _{C} + \vert 1\rangle _{B}\vert 1\rangle _{C})
\label{EPR4}
\end{eqnarray}
where $\vert 0\rangle $ and $\vert 1\rangle $\ are the
eigenvectors of the Pauli operator $\sigma _{z}$ (for example the
polarizations along the z direction). The subscripts $B$ and $C$
indicate the two correlated photons in each EPR pair. The four
local unitary operations $U_i$ $(i=0,1,2,3)$ can transform one of
the Bell states into another,
\begin{eqnarray}
U_{0}&&\equiv I=\vert 0\rangle \langle 0\vert + \vert 1\rangle
\langle 1\vert , \label{O0}\\
 U_{1}&&\equiv \sigma _{z}=\vert
0\rangle \langle 0\vert -\vert 1\rangle \langle 1\vert ,
\label{O1}
\\
U_{2}&&\equiv \sigma _{x}=\vert 0\rangle \langle 1\vert + \vert
1\rangle \langle 0\vert , \label{O2}\\
U_{3}&&\equiv i\sigma _{y}=\vert 0\rangle \langle 1\vert -\vert
1\rangle \langle 0\vert, \label{O3}
\end{eqnarray}%
where $I$ is the $2\times 2$ identity matrix and $\sigma_i$ are
the Pauli matrices. For example,
\begin{eqnarray}
&&I \otimes U _{0}\vert \psi^-\rangle=\vert \psi^-\rangle,
\,\,\,\,\,\,\,\,\,\, I \otimes U _{1}\vert \psi^-\rangle=\vert
\psi^+\rangle,\label{L1}\\
&&I \otimes U _{2}\vert \psi^-\rangle=\vert \phi^-\rangle,
\,\,\,\,\,\,\,\,\,\, I \otimes U _{3}\vert \psi^-\rangle=\vert
\phi^+\rangle.\label{L3}
\end{eqnarray}

Although the topological structure of a QSDC network can be loop or
star, similar to QKD network
\cite{Phoenix,Townsend,Biham,MUQKDguo,DLMXL,LZWD}, its subsystem can
be simplified to that in Fig.1. That is, a QSDC network is composed
of many subsystems (the small network cells) and there are three
roles in each cell, the server (Alice), the sender (Bob) and the
receiver (Carol). Alice provides the service for preparing the
quantum signal. Bob is the man who wants to send a message to Carol
privately. If Bob and Carol are not in the same branch on the
network \cite{MUQKDguo}, we assume that the server of the branch
with the sender Bob provides the service for preparing the quantum
signal, and the other servers provide the quantum line for Bob and
Carol (forbid all the others to use it) in a time slot \cite{DLMXL}.
Then the principle of this QSDC network is explicit if we describe
clearly the subsystem in Fig.1.

\begin{figure}[!h]
\begin{center}
\includegraphics[width=8cm,angle=0]{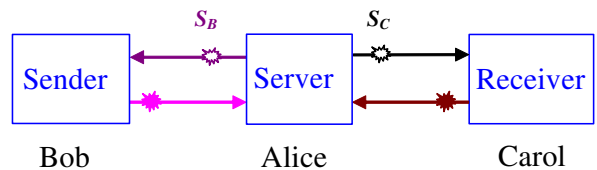} \label{f1}
\caption{ The subsystem of the present QSDC network. There are two
sequences of photons, $S_B$ and  $S_C$, which are transmitted to
the sender Bob and the receiver Carol, respectively.  The server,
Alice provides the service for preparing and measuring the Bell
states in the sequence of the EPR pairs. The legitimate users
exploit the four local unitary operations to encode their message
and complete the eavesdropping check with choosing one of the two
measuring bases $\sigma_z$ and $\sigma_x$ randomly.}
\end{center}
\end{figure}

Now, let us describe our bidirectional QSDC network scheme in
detail. First, we only consider the ideal condition. That is, we
assume that there is no noise and losses in the quantum channel, and
the devices are perfect. The case with a practical quantum line and
devices will be discussed in section IV. For the subsystem, the QSDC
can be implemented with seven steps.

(1) All the users on the network agree that the server, Alice
prepares an ordered $N$ EPR pairs in the same quantum state $
\vert \psi^- \rangle _{BC} =\frac{ 1}{\sqrt{2}}(\vert 0\rangle
_{B}\vert 1\rangle _{C} - \vert 1\rangle _{B}\vert 0\rangle _{C})$
for the quantum communication in each round. Alice divides it into
two sequences, $S_B$ and $S_C$. The $S_B$ is composed of all the
$B$ photons in the $N$ ordered EPR pairs, and the $S_C$ is
composed of all the $C$ photons.

(2) Alice sends the two sequences $S_B$ and $S_C$ to Bob and
Carol, respectively.

(3) After receiving the $S_B$ and $S_C$ sequences, Bob and Carol
choose randomly a sufficiently large subset of the $N$ EPR pairs as
the samples for checking eavesdropping,  and they measure each
photon in each sample EPR pair with the measuring basis (MB)
$\sigma_z$ or $\sigma_x$ chosen randomly. They complete the error
rate analysis by comparing the outcomes in public, same as that in
the Bennett-Brassard-Mermin 1992 (BBM92) QKD protocol \cite{bbm92}.
If the error rate is zero, they continue to next step, otherwise
they abandon their transmission and repeat the quantum communication
from the beginning.

For preventing the eavesdropper, the dishonest server from
eavesdropping by using Trojan horse attack with multi-photon signal
\cite{gisin} (For the attack with some invisible photons, the
parties can exploit a suitable filter to allow just the photons with
some a special frequency to be operated \cite{gisin}.), Bob and
Carol can complete the eavesdropping check with two steps
\cite{improving}. They can divide the samples into two parts. One is
used to determine whether there are many photons in the sample
signal. The other is used to determine whether the state of the
quantum signal is disturbed. For the first part, the two parties,
Bob and Carol can use a photon beam splitter to split the quantum
signal, and measure them with two single-photon detectors. If there
are many photons in the quantum signal received by the two parties,
the two detectors will both be clicked with a large probability for
each of the samples. In this way, the attack with Trojan horse can
be found out. For the other samples, Bob and Carol measure them with
two MBs $\sigma_z$ and $\sigma_x$ chosen randomly, same as that in
Ref. \cite{improving}.

For authenticating the outcomes published by Bob and Carol, they
should share a certain amount of classical key in advance, or they
have a special classical channel in which the information cannot be
altered even though any one can eavesdrop it, same as QKD
\cite{gisin}.

(4) Carol chooses randomly one of the four local unitary
operations $\{U_0, U_1, U_2, U_3\}$ which represent the two bits
of information 00, 11, 01 and 10 respectively, on each photon in
the $S_C$ sequence (except for the photons chosen for
eavesdropping check), and then she tells Bob the fact that she has
operated her sequence $S_C$. She sends the $S_C$ sequence to the
server Alice. The operations done by Carol is denoted as $U_C$.

(5) Bob encodes his message on the photons in the $S_B$ sequence
with one of the four local unitary operations $\{U_i\}$
($i=0,1,2,3$), say $U_B$, according to the message $M_B$,  and
then she also sends the $S_B$ sequence to the server Alice.

For analyzing the error rate of this transmission, Bob picks out
$k$ photons (the $k$ sample photon pairs are composed of them and
the correlated photons in the $S_C$ sequence) randomly
distributing in the $S_B$ sequence and performs one of the four
unitary operations randomly before he encodes the $S_B$ sequence.
The number $k$ is not big as long as it can provide an analysis
for the error rate.

(6) Alice performs Bell state measurements on  the photon pairs
and publishes the outcomes $U_A = U_B \otimes U_C$.

(7) Bob and Carol exploit the $k$ photon pairs chosen as the sample
pairs by Bob in advance to analyze the security of the whole quantum
communication and estimate its error rate. In detail, Bob tells
Carol her operations on the sample pairs and Carol compares them
with the outcomes published by Alice. If the error rate is zero,
Carol reads out the message $M_B$ with $U_B = U_A \otimes U_C$.
Otherwise, they discard the results.

In a QSDC network scheme, the most powerful eavesdropper may be
the dishonest server Alice as she prepares the quantum signal and
she can hide her eavesdropping with cheating besides other
strategies. If it is secure for the untrustworthy server, the
present QSDC network scheme is secure for any eavesdropper. So we
only discuss the eavesdropping done by the server Alice below.

The operations $U_C$ done by Carol are used to shield the effect of
the code done by Bob. It is equivalent to encrypting Bob's message
with an one-time pad crypto-system. The random key is just the
operations chosen randomly by Carol. In this way, the present QSDC
network is secure if the transmission of the two sequences from the
server Alice to the users Bob and Carol is secure as Alice's action
on the last stage can only obtain the combined outcome $U_A = U_B
\otimes U_C$ which will be published by the server and the cheat
that Alice publishes a wrong information in both the first stage
when the quantum signal is transmitted from Alice to the users and
the last stage will be found out by Bob and Carol with the
comparison of the outcomes of the $k$ sample photon pairs. The
transmission of the two sequences $S_B$ and $S_C$ from the server to
the users is similar BBM92 QKD protocol \cite{bbm92}. The difference
is just that the photons are transmitted in a quantum data block in
the present scheme, but one by one in the latter. The BBM92 QKD
protocol is proven secure in both  an ideal condition \cite{IRV} and
a practical condition \cite{WZY}. Hence, the present QSDC network
scheme can be made to be secure. Moreover, as almost all the
instances can be used to carry the message except for those for
eavesdropping check, the intrinsic efficiency for qubits $\eta_q$ in
the present QSDC network scheme approaches the maximal value 1.
\begin{eqnarray}
\eta_q\equiv \frac{q_u}{q_t},
\end{eqnarray}
where $q_u$ and $q_t$ are the qubits useful and total qubits for
the transmission. Each photon pair can carry two bits of message
which is the maximal source capacity for a two-photon entangled
state in quantum communication \cite{gisin,book}. The users
(except for the server) are required to have the capability of
single-photon measurements, which may make this network scheme
convenient in application.

\subsection{QSDC network with entanglement swapping}
In a practical quantum channel, there are noise and loss which maybe
affect the security of the network communication. The quantum
channel, a sequence of EPR pairs in the same state
$\vert\psi^-\rangle_{BC}=\frac{1}{\sqrt{2}}(\vert 0\rangle_B \vert
1\rangle_C - \vert 1\rangle_B \vert 0\rangle_C)$ can be set up
securely with entanglement purification \cite{ep,deutschqpa}. But
this case does not take place when the quantum states are  encoded
with the local operations. In this way, the users can accomplish the
quantum communication with entanglement swapping
\cite{entanglementswapping}. In detail, the subsystem of the QSDC
network can work as follows:

(1) The server Alice provides the service for Bob and Carol to
securely share a sequence of EPR pairs in the same state $\vert
\psi^-\rangle$, same as that in the bidirectional QSDC network
scheme discussed above.

(2) After purifying the EPR pairs \cite{ep}, Bob and Carol can
obtain a short sequence of maximally entangled two-photon states.
The two users divide them into some groups. There are two EPR pairs
in each group, say $\vert \psi^-\rangle_{B_1C_1}$ and $\vert
\psi^-\rangle_{B_2C_2}$. Bob encodes his message on the first EPR
pair in each group with the operations $U_i$ (i=0,1,2,3) and then
performs a Bell-basis measurement on the photons $B_1$ and $B_2$. He
announces his outcome in public.

(3) The receiver Carol takes a Bell-basis on his photons $C_1$ and
$C_2$ to read out the message. That is,
\begin{eqnarray}
\vert \psi^-\rangle_{B_1C_1}&\otimes & \vert \psi^-\rangle_{B_2C_2}
= \frac{1}{2}(\vert
\psi^-\rangle_{B_1B_2}\vert \psi^-\rangle_{C_1C_2}\nonumber\\
&& - \vert \psi^+\rangle_{B_1B_2}\vert \psi^+\rangle_{C_1C_2}
  -
\vert \phi^-\rangle_{B_1B_2}\otimes \vert \phi^-\rangle_{C_1C_2} \nonumber\\
&& + \vert \phi^+\rangle_{B_1B_2}\vert \phi^+\rangle_{C_1C_2}),\\
\vert \psi^+\rangle_{B_1C_1}&\otimes &\vert \psi^-\rangle_{B_2C_2} =
\frac{1}{2}(\vert
\psi^+\rangle_{B_1B_2}\vert \psi^-\rangle_{C_1C_2}\nonumber\\
&& - \vert \psi^-\rangle_{B_1B_2}\vert \psi^+\rangle_{C_1C_2}
  +
\vert \phi^+\rangle_{B_1B_2}\vert \phi^-\rangle_{C_1C_2} \nonumber\\
&& - \vert \phi^-\rangle_{B_1B_2}\vert \phi^+\rangle_{C_1C_2}),\\
\vert \phi^-\rangle_{B_1C_1}&\otimes &\vert \psi^-\rangle_{B_2C_2} =
\frac{1}{2}(\vert
\phi^-\rangle_{B_1B_2}\vert \psi^-\rangle_{C_1C_2}\nonumber\\
&& - \vert \phi^+\rangle_{B_1B_2}\vert \psi^+\rangle_{C_1C_2}
  -
\vert \psi^-\rangle_{B_1B_2}\vert \phi^-\rangle_{C_1C_2} \nonumber\\
&& + \vert \psi^+\rangle_{B_1B_2}\vert \phi^+\rangle_{C_1C_2}),\\
\vert \phi^+\rangle_{B_1C_1}&\otimes &\vert \psi^-\rangle_{B_2C_2} =
\frac{1}{2}(\vert
\phi^+\rangle_{B_1B_2}\vert \psi^-\rangle_{C_1C_2}\nonumber\\
&& - \vert \phi^-\rangle_{B_1B_2}\vert \psi^+\rangle_{C_1C_2}
  +
\vert \psi^+\rangle_{B_1B_2}\vert \phi^-\rangle_{C_1C_2} \nonumber\\
&& - \vert \psi^-\rangle_{B_1B_2}\vert \phi^+\rangle_{C_1C_2}).
\end{eqnarray}
Carol can deduce Bob's operation according to the result of her
Bell-state measurement and the information published by Bob easily.

In this QSDC network scheme, the sender Bob need not transmit the
qubits to the server Alice after he encoded his message on them.
That is, Bob and Carol do not give the chance for Alice to access
the qubits again when the quantum channel (a sequence of EPR pairs
shared) is confirmed to be secure. This principle will improve the
security of the quantum communication in a noise condition as the
qubits do not suffer from the noise and losses of the quantum line
again. The effect of the noise in quantum line on the qubits which
are transmitted between the server and the users can be eliminated
with entanglement purification \cite{ep} and quantum privacy
amplification \cite{deutschqpa}. Thus this network scheme can be
made to be secure. On the other hand, the users should have the
capability of taking a Bell-basis measurement on their qubits in
this QSDC network scheme, which will increase the difficulty of its
implementation in a practical application. Moreover, each EPR pair
shared between the sender and the receiver can carry only one bit of
information in theory, half of that in the bidirectional one.

\section{security analysis}

In essence, the server Alice first provides the service for the
users to share a sequence of EPR pairs in these two QSDC network
schemes, and the two authorized users confirm the security of the
quantum channel and then encode and decode the secret message after
they purified their EPR pairs. With the analysis for the samples by
using two MBs, $\sigma_z$ and $\sigma_x$, the sender and the
receiver can share securely a sequence of EPR pairs, similar to
BBM92 QKD protocol \cite{bbm92}. The difference is just that all the
qubits in BBM92 protocol are measured and the authorized users use
classical privacy amplification to distill a private string. In
these two QSDC network schemes, Bob and Carol can also use the
entanglement purification \cite{ep} and quantum privacy
amplification \cite{deutschqpa} to share a sequence of EPR pairs
securely. In this way, their security can be the same one. As the
security of the network schemes depend on completely that of the
quantum channel, they are secure since the quantum channel can be
made to be secure with entanglement purification
\cite{ep,deutschqpa}. Surely, the users can use the technique of
entanglement purity testing code \cite{testcode} to save the
entangled source largely for setting up their quantum channel.

We can also use the relation of the  error rate  and the
correspondent maximal amount of information obtainable for the
suspect server from a photon in each pair to demonstrate the
security of the quantum channel set up following the ideas in Refs.
\cite{bf,W1,D1,W2,D2,LM,fggnp,NG,ssz,sg}. In fact, the effect of the
eavesdropping on the two photons in an EPR pair with two unitary
operations is equal to that on one photon with another unitary
operation \cite{Preskill}, i.e., $(U_{eB}\otimes U_{eC})\vert
\psi^-\rangle_{BC} =(U_e\otimes I)\vert \psi^-\rangle_{BC}$. As Bob
and Carol check eavesdropping with choosing the two MBs $\sigma_z$
and $\sigma_x$ randomly, same as the BBM92 QKD protocol
\cite{bbm92}, the eavesdropping done by Eve can be realized by a
unitary operation, say, $\hat{E}$ on a larger Hilbert space. That
is, Eve can perform the unitary transformation $\hat{E}$ on the
photon $B$ and the ancilla $e$ \cite{NG,LM} whose state is initially
in $\vert 0\rangle$ \cite{Preskill}.
\begin{eqnarray}
\hat{E}\vert 0\rangle_{B} \vert 0\rangle &=& \sqrt{F}\vert 0\rangle
\vert e_{00}\rangle + \sqrt{D}\vert 1\rangle \vert
e_{01}\rangle, \label{ut3}\\
\hat{E}\vert 1\rangle_{B } \vert 0\rangle &=& \sqrt{D}\vert 0\rangle
\vert e_{10}\rangle + \sqrt{F}\vert 1\rangle \vert e_{11}\rangle
\end{eqnarray}
where $F$ is the fidelity of the state of the photon $B$ after the
eavesdropping, $D$ is the probability that Bob and Carol can detect
the action of eavesdropper, and the unitary of the operation
$\hat{E}$ requires the relations as follows
\begin{eqnarray}
\langle e_{00} \vert e_{00} \rangle + \langle e_{01} \vert e_{01}
\rangle =  F + D =1,
\\
\langle e_{10} \vert e_{10} \rangle + \langle
e_{11} \vert e_{11} \rangle =  D  + F =1,\\
\langle e_{00} \vert e_{10} \rangle + \langle e_{01} \vert e_{11}
\rangle=0.
\end{eqnarray}

With the unitary operation $\hat{E}$ done by Eve and the unitary
operations $U_i$ done by Bob with the probabilities $P_i$
$(i=0,1,2,3)$, the final state of the photon $B$ and the ancilla
$e$ is described as follows \cite{D2}.
\begin{eqnarray}
\varepsilon(\rho_{B})=\sum_{i=0}^3 P_i\varepsilon_{U_i}(\rho_{B}),
\end{eqnarray}
where $P_i$ is the probability encoded with the operation $U_i$,
$\rho_{B}=Tr_C(\rho_{BC})=\frac{1}{2}\left(
\begin{array}{cc}
1 & 0 \\
0 & 1%
\end{array}%
\right)$, and $\varepsilon_{U_i}$ is quantum operation describing
the evolution of the initial state $\rho_{B}$, i.e.,
\begin{eqnarray}
\varepsilon_{U_i}(\rho_{B})=U_i\hat{E}\rho_{B}\otimes\vert
e\rangle\langle e\vert \hat{E}^+U_i.
\end{eqnarray}
The accessible information extracted from the state
$\varepsilon(\rho_{B})$ is no more than the Holevo bound
\cite{book,D2,bf}, i.e.,
\begin{eqnarray}
I_{C}\leq S(\varepsilon(\rho_{B}))-\sum_{i=0}^3 P_i
S(\varepsilon_{U_i}(\rho_{B}))\equiv I_0,
\end{eqnarray}
where $S(\rho)$ is the von-Neumann entropy of the state $\rho$,
i.e.,
\begin{equation}
S(\rho)=-Tr\{\rho log_2 \rho\}=\sum\limits_{i=0}^{3}-\lambda
_{i}\log _{2}\lambda _{i}, \label{information1}
\end{equation}
where $\lambda_i$ are the roots of the characteristic polynomial
det$(\rho-\lambda I)$ \cite{bf}. As we use a mixed state to describe
the photon $B$, Eve can steal the bit value or the phase of the
operations done by Bob, same as Ref. \cite{D2}. In fact, the
information that Eve can steal is no more than twice of $I_0$.

Certainly, Eve can eavesdrop the operation done by Bob $U_B$ with
two unitary operations and two ancilla systems, but it does not
increase the information about the operation $U_B$ and not decrease
the probability $D$. Now, let us calculate the value of $I_0$ in the
case that Carol chooses the four local unitary operations with the
same probability $P_i=\frac{1}{4}$. Define an orthonormal base
$\{\vert 00\rangle$, $\vert 01\rangle$, $\vert 10\rangle$, $\vert
11\rangle \}$ which spans the generic subspace of the Hilbert space
$H_{C} \otimes H_e$ support of $\varepsilon(\hat{\rho}_{C})$. With
the Eqs. (\ref{ut3})-(\ref{information1}), we can obtain the
relation between $I_0$ and the probability $D$ as following.
\begin{eqnarray}
I_0 &=& -Dlog_2 D - (1-D)log_2(1-D).
\end{eqnarray}
When $D=0.25$, the information that Eve can obtain $I_E\leq
2I_0=1.62<2$. That is, Bob and Carol can set up the quantum channel
securely with entanglement purification \cite{ep} and quantum
privacy amplification \cite{deutschqpa}, similar to QKD with a
classical privacy amplification \cite{book}.

In a practical quantum channel, there are noise and losses which
will threaten the security of quantum communication. The present
QSDC network schemes are secure in a closely ideal condition, but it
is also affected by the noise and the losses in a practical channel,
same as the two-step QSDC protocol \cite{two-step} and others
\cite{QOTP,Wangc,Wangc2,bf,cai,Nguyen,mancpl,caiA,Lucamaini}. The
quantum states are transmitted in a quantum data in the present QSDC
network scheme, which will ensure it to overcome the effect of the
noise in the practical channel as the parties can use quantum
privacy amplification technique \cite{deutschqpa} to improve the
security of the quantum states transmitted. This advantage happens
only in the QSDC schemes with quantum data block, such as Refs.
\cite{two-step,QOTP,Wangc}. In order to reduce the effect of the
losses, another quantum technique, quantum teleportation \cite{bw}
can be used to determine whether the receiver has obtained the
photons sent by the server Alice in the process of the transmission
for  the $S_C$ sequence, same as that in the two-step QSDC scheme
\cite{two-step}. That is, Carol should prepare another $N$ EPR pairs
and perform quantum teleportation on the photons in the $S_C$
sequence with her EPR pairs. If the teleportation successes, she
tells Bob to encode the correlated photon in the $S_B$ sequence. On
the other hand, the present network schemes can be also used to
distribute a private key and the users need not exploit quantum
teleportation to improve its security in a loss channel.

\section{Discussion and summary}

It is of interest to point out that the stream-QSDC schemes
\cite{bf,cai,caiA,Nguyen,mancpl,Lucamaini} do not work for a network
as they are only quasi-secure in a practical channel. It is
difficult for the users to do the error correction and the privacy
amplification in those QSDC schemes
\cite{bf,cai,caiA,Nguyen,mancpl,Lucamaini}. In particular, the
privacy amplification cannot be accomplished as the photon is
transmitted one by one and the information transmitted is the
deterministic message, not a random key. The QSDC schemes in Refs.
\cite{QOTP,Wangc} cannot be used to complete the task of a QSDC
network in a simple way as the server who prepares the quantum
signal can steal almost all the information about the message
without being found out in a noise channel. For example, in the QSDC
network with the quantum one-time pad scheme \cite{QOTP}, the server
can intercept the photons encoded by Bob and read out the operations
freely. Certainly, the two users can exploit some other technique to
improve the security, but the classical information exchanged will
increase largely, same as the QSDC schemes with quantum teleporation
\cite{yan} and quantum swapping \cite{zhangzj} which are close to
QKD.

In summary, we have proposed two  QSDC network schemes with EPR
pairs. In the first one, the server prepares and measures the EPR
pairs and the users exploit the four local unitary operations to
encode their message, which makes the users on the network more
convenient than others in a practical application. One can
communicate any other one on the network securely as they can
perform a quantum privacy amplification on the quantum states
transmitted in a noise channel. Its intrinsic efficiency for qubits
and source capacity are both high as almost all of the instances are
useful and each EPR pair can carry two bits of information. In the
second QSDC network scheme, the users exploit entanglement swapping
to transmit the secret message, which will improve its security in a
noise quantum line at the risk of lowering its source capacity and
increasing the difficulty of the experimental implementation for the
users. Also, the four requirements for a real secure point-to-point
quantum direct communication scheme are discussed in detail and the
present QSDC network schemes satisfy all the four requirements.

\section*{ACKNOWLEDGEMENTS}
This work is supported by the National Natural Science Foundation
of China under Grant Nos. 10447106, 10435020, 10254002, A0325401
and 10374010, and Beijing Education Committee under Grant No.
XK100270454.

\end{document}